\documentclass[aps,prl]{revtex4}
\usepackage{graphicx}
\begin{document}

\title{Anharmonicity and self-similarity of the free energy landscape
  of protein G} \author{F. Pontiggia$^1$, G. Colombo$^2$,
  C. Micheletti$^1$ and H. Orland$^3$} 

\affiliation{ \centerline{$^1$ International School for Advanced
Studies and CNR-INFM Democritos, Via Beirut 2-4, 34014 Trieste,
Italy} \centerline{$^2$ Consiglio Nazionale delle Richerche, Via M.
Bianco 9, 20131 Milano, Italy} \centerline{$^3$ Service de Physique
Theorique, CE-Saclay, CEA 91191 Gif-sur-Yvette Cedex, France} }

\date{\today}
\begin{abstract}
The near-native free energy landscape of protein G is investigated through
0.4$\mu$s-long atomistic molecular dynamics simulations in explicit
solvent. A theoretical and computational framework is used to assess
the time-dependence of salient thermodynamical features. While the
quasi-harmonic character of the free energy is found to degrade in a
few ns, the slow modes display a very mild dependence on the
trajectory duration. This property originates from a striking
self-similarity of the free energy landscape embodied by the
consistency of the principal directions of the local minima, where the
system dwells for several ns, and of the virtual jumps connecting
them.
\end{abstract}
\maketitle

The multiscale protein dynamics observed in single-molecule
experiments \cite{Xie1,BaldiniScience} provides a vivid reverberation
of the complexity of the free energy surface of folded proteins
\cite{substates,Funnel2}. Complementing these phenomenological
observations with detailed theoretical and computational
investigations has proved challenging. For instance, most of the
atomistic molecular dynamics (MD) explorations of the free-energy
landscape organization have probed features with a time resolution
below the ns scale. Though limited, this time span is sufficient to
reveal the presence, within the native basin, of a plethora of local
energy minima corresponding to minute structural differences
\cite{McCammon.BIOP.1984,BrooksIII.JCC.1995,GO.PROT.1998}. Within the
short time spent in each of these minima, about 1 ps
\cite{BrooksIII.JCC.1995}, the normal modes analysis based on the
local quadratic approximation of the energy adequately captures the
system dynamics\cite{Tirion.PRL.1996}. It is surprising, however, that
the 10-20 slowest normal modes of any of these minima are sufficient
to describe the large scale conformational changes occurring over tens
of ns or more e.g. switches between active/inactive enzymatic forms
etc.  \cite{enzymes,ENM}.  This remarkable property, exhibited by
generic globular proteins \cite{enzymes}, indicates a special
multiscale organization of the free energy landscape. In this study we
dissect, by quantitative means, this organization for a particular
globular biopolymer, protein G. In particular, we analyse the
properties of the phase space explored by atomistic molecular dynamics
simulations covering as much as 0.4 $\mu$s. This represents an
increase of more than two orders of magnitude in duration with respect
to previous related investigations and hence allows an unprecedented
insight into the organization of the free energy surface
\cite{BrooksIII.JCC.1995,GO.PROT.1998}. The latter is shown to
comprise minima whose self-similarity extends, unexpectedly, to the
largest examined time scales. The observed self-similarity is of a
peculiar kind as it pertains only to the principal directions of the
minima which otherwise differ significantly in depth and breadth. The
findings provide a novel insight on the robustness of the large-scale
concerted movements in proteins obtained either through principal
component analysis of MD simulations or from simplified mesoscopic
models.

We first consider the assumption, at the basis of several elastic
network approaches \cite{bahar97,ENM}, that the free energy governing
the near-native dynamics of a protein has a quadratic character:
\begin{equation}
{\cal F} = \sum_{i,j,\mu,\nu} \ \ M_{ij,\mu\nu}\, \delta
r_{i,\mu}\,\delta r_{j,\nu}
\label{eqn:vexp}
\end{equation}
\noindent where $\delta r_{i,\mu}$ indicates the $\mu$th Cartesian
component of the vector displacement of the $i$th amino acid
(represented by the C$_\alpha$ atom) from its native reference
position. Despite the minimalistic character, harmonic models based on
eq. (\ref{eqn:vexp}) have proved capable of capturing large-scale
movements occurring over time scales that vastly exceed those of
expected validity for local quadratic expansions of the free energy
\cite{BrooksIII.JCC.1995,GO.PROT.1998}.
Prompted by these results we investigate in detail the extent to which
extensive atomistic MD simulations support the simple quadratic form
of eqn. (\ref{eqn:vexp}). First we shall recover the
matrix of effective couplings, $M$, that best describes one or more
constant-temperature dynamical trajectories (from which
roto-translations are previously removed by optimally superimposing
each recorded system configuration to a given reference structure.)
The effective matrix $M$ associated to an equilibrated trajectory of
duration $\Delta t$ will be obtained from the covariance matrix, $C$
\cite{garcia92,Amadei.PROT.1993}:
\begin{equation}
{C}_{ij,\mu\nu}(\Delta t) = \langle [ r_{i,\mu}(t) -
\langle {r}_{i,\mu} \rangle ] \ [r_{j,\nu}(t) - \langle r_{i,\nu}\rangle ]\rangle
\label{eqn:cij}
\end{equation}
\noindent where $\vec{r}_{i}(t)$ is the position of the $i$th amino
acid at time $t$ and the brackets denote the time-average over a
specific time interval of duration $\Delta t$.  Within the quadratic
approximation, the coupling matrix is given by $M(\Delta t) =
\tilde{C}^{-1}(\Delta t)/\kappa_B\, T$, where $\kappa_B T$ is the
thermal energy and the tilde indicates the removal of the six
roto-translational zero eigenspaces of $C$ prior to inversion.

Next, by considering longer and longer trajectories, we shall examine
how $M$ depends on the breadth of the visited phase space. The
consistency of two $M$ matrices will be assessed by comparing both the
corresponding entries and their ten slowest modes. The latter
correspond to the eigenvectors of $M$ associated to the smallest
non-zero eigenvalues. The similarity of two sets of slow modes,
$\{v\}$ and $\{w\}$, is measured, as customary, in terms of the root
mean square inner product (RMSIP):
\begin{equation}
RMSIP = \sqrt{\sum_{i,j=1,...,10} | \vec{v}_i \cdot \vec{w}_j| ^2 /10} .
\label{eqn:rmsip}
\end{equation}

\begin{figure}[t!]
\includegraphics*[width=5.4in]{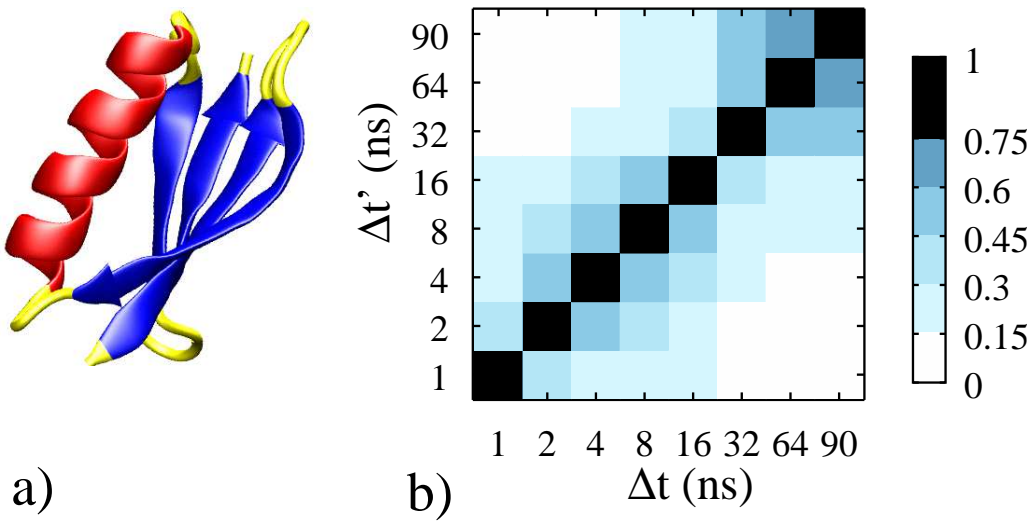}
\caption{[Color online] (a) Ribbon representation of
    protein G. (b) Color-coded plot of the Kendall's correlation
    coefficient, $\tau$, between corresponding elements of the
    coupling matrices, $M$, calculated over the first $\Delta t$ and
    $\Delta t^\prime$ ns of the first trajectory. Matrix elements
    along the diagonal or pertaining to consecutive $C_\alpha$'s were
    omitted in the calculation of $\tau$.}
\label{fig:correlations}
\end{figure}

\noindent This analysis is undertaken for protein G, an important
signalling biomolecule shown in Fig. \ref{fig:correlations}a. It
possesses a non-trivial $\alpha/\beta$ tertiary organization despite
its moderate length, 56 residues, which makes it amenable to extensive
MD simulations. The 1PGB crystal structure was taken as the starting
point of 4 MD runs in explicit solvent each of 100 ns. First the
system was energy minimized after solvation with the single-point
charge water model in an octahedral box containing an explicit solvent
layer of 1.2 nm. The density was adjusted by a 100ps-long MD in NPT
conditions by weak coupling to a bath of constant pressure (P = 1 bar,
coupling time $\tau$=0.5 ps). Subsequently the four different 100-ns
long trajectories were started with different sets of Maxwellian
(T=300 K) initial atomic velocities and integrated with the GROMACS
engine\cite{GROMACS_3.0_01} The particle mesh Ewald method was used
for electrostatics and thermostatting was provided by a thermal bath
(coupling time $\tau$=0.1ps). To remove correlations due to the same
starting structure we omitted the initial 10 ns from the
analysis. The energy-mininised initial structure is,
on the other hand, used as the common reference configuration to
roto-translate, and hence align in space, all four trajectories. From
the remaining 90 ns long trajectories, intervals of increasing
duration $\Delta t$ are used to compute the coupling matrices
$M(\Delta t)$ from inversion of the covariance ones. 

Pairs of $M$ matrices from the different runs and/or for increasing
$\Delta t$ are first compared by measuring the Kendall's correlation
coefficient\cite{NR}, $\tau$, of the $\sim$13000 corresponding
distinct entries.  We employ Kendall's $\tau$ as it provides a robust
measure of data association with no prior assumption of the parametric
dependence of the examined data sets. To convey the
degree of consistency of $M$ upon extending the length of the
simulation we portray in Fig. \ref{fig:correlations}b the Kendall's
$\tau$ pertaining to pairs of intervals of different durations from
the first trajectory. As visible in the figure, extending the duration
of the trajectory, be it 1 or 16 ns, by four or more times leads to a
substantial deviation of corresponding entries of $M$ (with $\tau \sim
0.30$). Consistently, halving each trajectory and considering the
correlation among the two halves gives values of $\tau$ between 0.17
and 0.25. The consistency of $M$ matrices of different trajectories
is, furthermore, much poorer and almost independent on the compared
time spans. In fact, pairwise comparisons of different runs over the
first ns or over the entire 90-ns duration trajectories yield values
of $\tau$ in the [0.05- 0.15] range.

These values indicate a substantial degree of heterogeneity in
corresponding matrix entries and hence point to the impossibility of
having a robust definition of $M$ even over hundreds of ns. An
analogous loss of consistency is observed in the covariance matrices,
with $\tau \sim 0.15$ for pairwise comparions of corresponding $C$
entries calculated over the entire trajectories.  Yet, despite the
fact that no asymptotic value appears to be reached by the $M$ (or
$C$) entries, the latter display general properties which are robust
against $\Delta t$ and hence may be taken into account to improve
elastic network models based on the {\em a priori} quadratic
approximation of eq. \ref{eqn:vexp}. For example, consistently with
previous findings based on the Hessian analysis
\cite{Hinsen.CHEMP.2000}, consecutive $C_\alpha$'s are coupled by
interactions that are more than 10 times stronger than non-consecutive
ones. Furthermore, in all trajectories, negligible pairwise couplings
are typically found among centroids at a distance larger than $8-9$
\AA. Interestingly, below this cutoff, pairwise interactions can have
either attractive or repulsive character (yet all eigenvalues of ${M}$
are clearly non-negative).

\begin{figure}[t!]
\includegraphics*[width=5.5in]{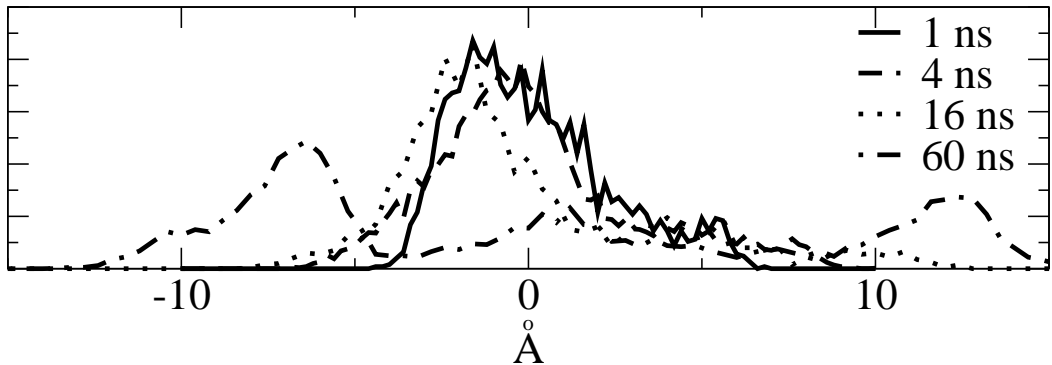}
\caption{Normalised distribution of the projection
  along the first slow mode of the conformational fluctuations
  encountered in intervals of increasing length from the first
  trajectory.}
\label{fig:projections}
\end{figure}

\begin{figure}[t!]
\includegraphics*[width=5.5in]{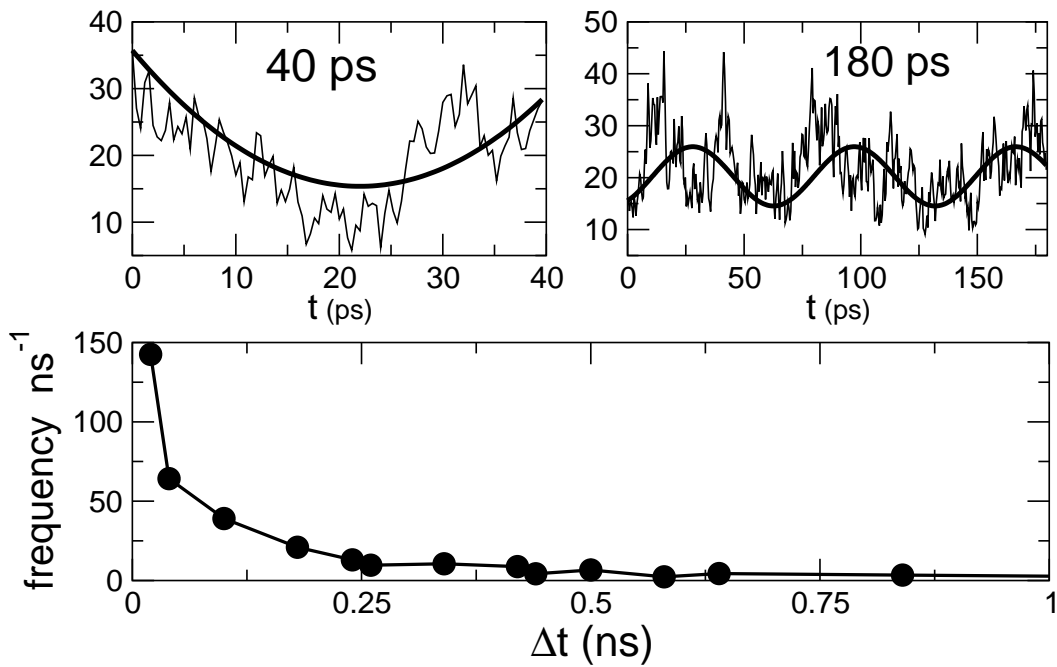}
\caption{Top: time evolution of the projection along the 1st
eigenvector and of the variational deterministic oscillator for two
intervals of different duration. Bottom: average value of the
variational oscillator frequency as a function of the interval
duration, $\Delta t$.}
\label{fig:opt_osc}
\end{figure}

We shall now address the origin of the apparent lack of robustness of
$M$ by considering, in place of the individual matrix entries, the
time dependence of the slow modes. For a system governed by a purely
quadratic free energy, the fluctuations around the average structure
have a Gaussian probability distribution along any of the eigenvectors
of $M$. The width of the Gaussian is largest for the slowest mode
which, corresponding to the direction of least curvature of ${\cal
F}$, mostly accounts for the system fluctuations in thermal
equilibrium. A valuable insight into the effective free energy
landscape described by $M(\Delta t)$ is hence obtained by considering
the distribution of the projections along the slowest mode of the
conformational fluctuations in trajectories of increasing duration.
Typical results are shown in Fig. \ref{fig:projections}.
\noindent It is noticed that, for $\Delta t = $ 1 ns, the
distributions have a unimodal character with a fair degree of
Gaussianity. In fact, for all four trajectories, the normalised
kurtosis $\kappa = (\langle x^4 \rangle - 3 \langle
x^2\rangle^2)/\langle x^4 \rangle$ ($x$ being the projection)
typically has values of 0.15 for $\Delta t$ up to 2 ns. For time spans
larger than 5-10 ns deviations from Gaussianity become very apparent
through the non-unimodal character of the distribution. The analysis
of the kurtosis in the fluctuations of individual amino acids reveals
that the largest deviations are observed for exposed regions,
particularly residues 30--40. Also, the eigenvalue of $M$ associated
to the slowest mode decreases as $\Delta t$ is increased, thus
suggesting a progressive weakening of the quadratic potential upon the
widening of the explored phase space. This softening can be vividly
illustrated from a novel perspective by means of a simplified, but
transparent, dynamical variational approach. More precisely, we shall
consider the projection of a trajectory along its slowest mode ${\vec
v}_1$and attempt to describe its time-varying amplitude with a
deterministic harmonic vibration of the amino acids:
\begin{equation}
\vec{r}^m_i(t) = \vec{r}^0_i(t=0) + a\, {\vec v}_{1,i} [\cos(\omega t + \phi) - \cos(\phi)]
\end{equation}
\noindent where the superscript $m$ is used to denote the coordinates
of the model oscillators which coincide with those of the real system
(superscript 0) at $t=0$. The quantities $\omega$, $\phi$ and $a$ are
obtained from a dynamical variational criterion
\cite{Pitard.EurPhysLett.1998}, namely by minimizing the time-averaged
total square deviation of the model trajectory from the real one,
$\langle\sum_i |\vec{x}^m_i(t) -\vec{x}^0_i(t)|^2 \rangle$.
As visible in Fig. \ref{fig:opt_osc}, for time spans of up to
fractions of a ns the oscillator is able to account satisfactorily for
the evolution of the true trajectory (manifestly overdamped over
$\Delta t >~ 0.5$ ns). The lower panel of Fig. \ref{fig:opt_osc},
presenting a visible decrease of the optimal frequency $\omega$,
illustrates how rapidly the curvature of the effective quadratic free
energy decreases as a function of $\Delta t$.  The softening reflects
the complexity and anharmonicity of the free energy landscape which,
as confirmed by the multimodal character of the distributions of
Fig. \ref{fig:projections}, is constituted by broad minima of varying
depth which are progressively explored as the dynamics advances.

\begin{figure}[t!]
\includegraphics*[width=5.5in]{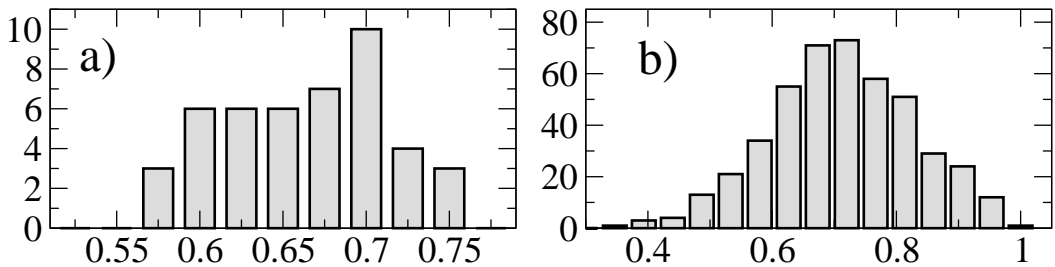}
\caption{(a) Histogram of the RMSIP values calculated for the
principal components of all pairs of top structural clusters. (b)
Histogram of the norm of the projection of all pairwise distance
vectors between cluster representatives along the top 10 principal
components of the largest cluster.}
\label{fig:clusters}
\end{figure}

\begin{figure}[t!]
\includegraphics*[width=5.0in]{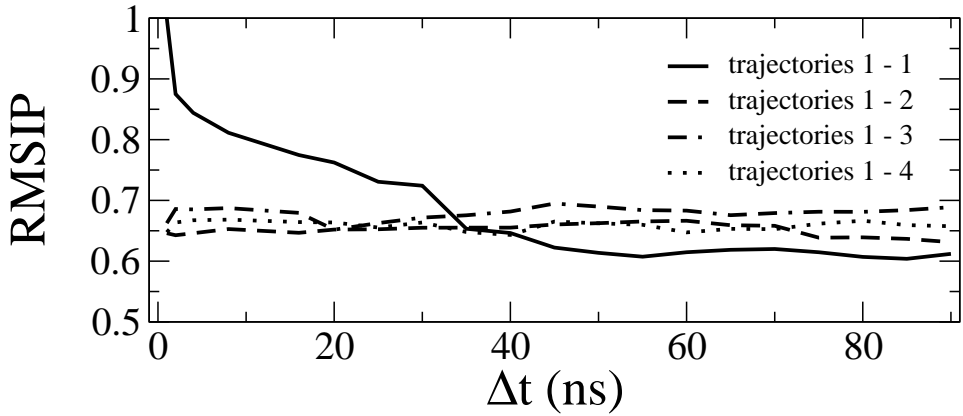}
\caption{RMSIP between the top essential dynamical spaces of the 1st
ns of trajectory 1 and intervals of longer duration, $\Delta t$, from
the same and other trajectories.}
\label{fig:rmsip}
\end{figure}

The properties of the most pronounced minima have been
examined in detail and lead to discovering that the self-similarity of
the free energy landscape, previously ascertained with a sub-ns time
resolution \cite{GO.PROT.1998}, extends to an unsuspectedly large
scale.  To characterize the salient minima we first performed a
cluster analysis of the structures generated in all four trajectories
\cite{Daura.ANGCHEM.1999} and considered the 10 most populated
sets. Each cluster, containing conformations at less than 1.5 \AA\
RMSD from the representative, gathers configurations from
time-intervals of very different length, from 1 to tens of ns,
originating from one to three of the four trajectories. By carrying
out a structural covariance analysis (with the time average of
eqn. \ref{eqn:cij} being replaced with an average over cluster
members) we identified the 10 principal components describing the
largest conformational changes in each cluster. By means of the RMSIP
we finally compared the 10 principal components of all distinct pairs
of clusters. The resulting distribution of RMSIP values indicated a
very strong consistency of the sets of principal directions, see
Fig. \ref{fig:clusters}a. Indeed, for the considered protein length,
numerical results indicate that if the set of principal components
were completely unrelated, the expected RMSIP value would be
$0.24\pm0.02$.  The distribution of values in Fig. \ref{fig:clusters}
is sufficiently distant from this random reference value to convey the
significance of the observed consistency of the principal components
of different clusters.  Strikingly, it was also found that the
difference vectors, $\vec{d}_{ij}$ connecting the representatives of
any pair of different clusters $i$ and $j$ are also well described by
the principal components of any of the clusters. For example, as shown
in Fig. \ref{fig:clusters}b, the top 10 principal components of the
largest cluster are usually sufficient to account for most of the norm
of the ``virtual jumps'' connecting the representatives of any two top
clusters. These facts reveal an appealing self-similar structure of
the free-energy profile: not only the deep free-energy minima
corresponding to the main clusters have similar principal components,
but also the virtual ``jumps'' connecting their representatives are
describable in the same low-dimensional space. This remarkable
property can be formalised by using a generalization of the covariance
decomposition of ref. \cite{GO.PROT.1998}. In fact, for a generic MD
trajectory that has visited several free-energy minima (clusters) the
covariance matrix reflects the structural heterogeneity within and
across the clusters \cite{GO.PROT.1998}.  More precisely:
\begin{equation}
{C} = \sum_l  w_l \{ { C}^l + \sum_{k,m} w_m \, w_k\, | \vec{d}_{l,k} \rangle
\langle \vec{d}_{l,m} | \}
\end{equation}
\noindent where $w_l$ is the weight of the $l$th cluster, that is the
fraction of time spent by the system in it, ${C}^l$ is the
covariance matrix of the $l$th cluster, and $\vec{d}_{l,m}$ is the
distance vector of the representative (average) structures of
clusters $l$ and $m$. We now build on the previous observation that a
single set of principal components $\vec{v}_1$,... $\vec{v}_n$ can
well describe both the essential spaces of any cluster $l$ and also
the inter-cluster distance vectors: ${\bf C}^l = \sum_n \lambda^l_n |
\vec{v}_n \rangle \langle \vec{v}_n | $ and $\vec{d}_{l,m} = \sum_n
c^{l,m}_n \vec{v}_n$. According to this approximation, the principal
eigenvectors of ${C}$, and hence the slow modes of $M$, would coincide
with $\vec{v}_1$,... $\vec{v}_n$ and thus remain unchanged over all
time scales. On the contrary, the associated eigenvalues would
explicitly depend on the MD duration through the time-dependent number
and weight of the visited clusters.
To verify these conclusions we have compared the essential spaces of
different intervals from the simulated trajectories. The results are
illustrated in Fig. \ref{fig:rmsip} which portrays the RMSIP
calculated between the essential spaces of the 1st ns for trajectory 1
with larger and larger time spans for the same and other trajectories.
It is seen that the top slowest modes are very robust against
increasing $\Delta t$ and remain consistent even augmenting the
simulation time by two orders of magnitude (from 1 to 90 ns). The
statistical significance of this result is highlighted by the
difference of RMSIP ranges in Fig. \ref{fig:rmsip} from the
aforementioned random reference value of 0.24. As a further comparison
we also considered the RMSIP value calculated over all pairs of 10
mid-ranking eigenvectors of the last 1 ns of all four
trajectories. Also this more stringent test indicates that the RMSIP's
in Fig. \ref{fig:rmsip} exceed the control value by at least 4
standard deviations and hence have a high statistical significance.

In summary, the analysis of the dynamics of protein G, a typical
globular protein, revealed an unexpectedly simple self-similar
structure of the various free energy minima and of the virtual jumps
connecting them. This remarkable feature reflects into the exceptional
robustness of the essential dynamical spaces (slow modes) calculated
over trajectories with very different duration. Instead, the typical
amplitudes projected along the slowest modes depends on the number of
visited minima as well as on their depth. As a result, the dynamical
projections have a strong dependence on the duration of the
simulation, a fact that accounts for the observed inconsistency of the
coupling (or covariance) matrix entries. The observed properties,
besides elucidating general features of the free energy landscape of
 one particular protein, have important practical
ramifications. In particular they provide a first perspective, for
understanding the scope and viability of coarse-grained elastic
network models as well as short MD simulations. Accordingly, it is
expected that the directionality of the slow modes/essential dynamical
spaces can be determined with considerable more confidence than the
amplitude of the associated dynamical
projections. These considerations provide a strong
motivation for investigating the validity and transferability of the
present analysis to other protein contexts.

We acknowledge support from FIRB 2003 (grant RBNE03B8KK) from the 
Italian Ministry for Education.

%%%%%%%%%%%%%%%%%%%%%%%%%%%%%%%%%%%%%%%%%%%%%%%%%%%%
%%\bibliography{./bibliografia/biblio}

\begin{thebibliography}{19}
\expandafter\ifx\csname natexlab\endcsname\relax\def\natexlab#1{#1}\fi
\expandafter\ifx\csname bibnamefont\endcsname\relax
  \def\bibnamefont#1{#1}\fi
\expandafter\ifx\csname bibfnamefont\endcsname\relax
  \def\bibfnamefont#1{#1}\fi
\expandafter\ifx\csname citenamefont\endcsname\relax
  \def\citenamefont#1{#1}\fi
\expandafter\ifx\csname url\endcsname\relax
  \def\url#1{\texttt{#1}}\fi
\expandafter\ifx\csname urlprefix\endcsname\relax\def\urlprefix{URL }\fi
\providecommand{\bibinfo}[2]{#2}
\providecommand{\eprint}[2][]{\url{#2}}

\bibitem[{\citenamefont{Min et~al.}(2005)\citenamefont{Min, English, Luo,
  Cherayil, Kuo, and Xie}}]{Xie1}
\bibinfo{author}{\bibfnamefont{W.}~\bibnamefont{Min}} {\em et al.}
  , \bibinfo{journal}{Acc.
  Chem. Res.} \textbf{\bibinfo{volume}{38}}, \bibinfo{pages}{923}
  (\bibinfo{year}{2005}).

\bibitem[{\citenamefont{Cannone et~al.}(2005)\citenamefont{Cannone, Chirico,
  and Baldini}}]{BaldiniScience}
\bibinfo{author}{\bibfnamefont{F.}~\bibnamefont{Cannone}},
  \bibinfo{author}{\bibfnamefont{G.}~\bibnamefont{Chirico}}, \bibnamefont{and}
  \bibinfo{author}{\bibfnamefont{G.}~\bibnamefont{Baldini}},
  \bibinfo{journal}{Science} \textbf{\bibinfo{volume}{309}},
  \bibinfo{pages}{1096} (\bibinfo{year}{2005}).

\bibitem[{\citenamefont{Frauenfelder et~al.}(1991)\citenamefont{Frauenfelder,
  Siglar, and Young}}]{substates}
\bibinfo{author}{\bibfnamefont{H.}~\bibnamefont{Frauenfelder}},
  \bibinfo{author}{\bibfnamefont{H.}~\bibnamefont{Siglar}}, \bibnamefont{and}
  \bibinfo{author}{\bibfnamefont{R.~D.} \bibnamefont{Young}},
  \bibinfo{journal}{Science} \textbf{\bibinfo{volume}{254}},
  \bibinfo{pages}{1598} (\bibinfo{year}{1991}).

\bibitem[{\citenamefont{Wolynes et~al.}(1995)\citenamefont{Wolynes, Onuchic,
  and Thirumalai}}]{Funnel2}
\bibinfo{author}{\bibfnamefont{P.~G.} \bibnamefont{Wolynes}},
  \bibinfo{author}{\bibfnamefont{J.~N.} \bibnamefont{Onuchic}},
  \bibnamefont{and}
  \bibinfo{author}{\bibfnamefont{D.}~\bibnamefont{Thirumalai}},
  \bibinfo{journal}{Science} \textbf{\bibinfo{volume}{267}},
  \bibinfo{pages}{1619} (\bibinfo{year}{1995}).

\bibitem[{\citenamefont{Levy et~al.}(1984)\citenamefont{Levy, Srinivasan,
  Olson, and McCammon}}]{McCammon.BIOP.1984}
\bibinfo{author}{\bibfnamefont{R.~M.} \bibnamefont{Levy}} {\em et al.} 
  \bibinfo{journal}{Biopolymers} \textbf{\bibinfo{volume}{23}},
  \bibinfo{pages}{1099} (\bibinfo{year}{1984}).

\bibitem[{\citenamefont{Janezic et~al.}(1995)\citenamefont{Janezic, Venable,
  and Brooks}}]{BrooksIII.JCC.1995}
\bibinfo{author}{\bibfnamefont{D.}~\bibnamefont{Janezic}},
  \bibinfo{author}{\bibfnamefont{R.}~\bibnamefont{Venable}}, \bibnamefont{and}
  \bibinfo{author}{\bibfnamefont{B.~R.} \bibnamefont{Brooks}},
  \bibinfo{journal}{J. Comp. Chem.} \textbf{\bibinfo{volume}{16}},
  \bibinfo{pages}{1554} (\bibinfo{year}{1995}).

\bibitem[{\citenamefont{Kitao et~al.}(1998)\citenamefont{Kitao, Hayward, and
  Go}}]{GO.PROT.1998}
\bibinfo{author}{\bibfnamefont{A.}~\bibnamefont{Kitao}},
  \bibinfo{author}{\bibfnamefont{S.}~\bibnamefont{Hayward}}, \bibnamefont{and}
  \bibinfo{author}{\bibfnamefont{N.}~\bibnamefont{Go}},
  \bibinfo{journal}{Proteins} \textbf{\bibinfo{volume}{33}},
  \bibinfo{pages}{496} (\bibinfo{year}{1998}).

\bibitem[{\citenamefont{Tirion}(1996)}]{Tirion.PRL.1996}
\bibinfo{author}{\bibfnamefont{M.~M.} \bibnamefont{Tirion}},
  \bibinfo{journal}{Phys. Rev. Lett.} \textbf{\bibinfo{volume}{77}},
  \bibinfo{pages}{1905} (\bibinfo{year}{1996}).


\bibitem{enzymes} W.G. Krebs {\em et al.}, Proteins {\bf 48}, 682
  (2002); A.L. Perryman, J.-H. Lin, and J.A. McCammon,
  Prot. Sci. {\bf 13}, 1108 (2004); D. Ming and M.E. Wall, Phys. Rev. Lett. {\bf 95},
  198103 (2005)


\bibitem{ENM} K. Hinsen , Proteins {\bf 33} , 417 (1998) ;
   A. R. Atilgan  {\em et al.} , Biophys. J. {\bf 80} , 505 (2001);
   M. Delarue adn Y-H Sanejouand , J. Mol. Biol. {\bf 320} , 1011
   (2002) ; C. Micheletti , P. Carloni , and A. Maritan , Proteins
   {\bf 55} , 635 (2004).

\bibitem[{\citenamefont{Bahar et~al.}(1997)\citenamefont{Bahar, Atilgan, and
  Erman}}]{bahar97}
\bibinfo{author}{\bibfnamefont{I.}~\bibnamefont{Bahar}},
  \bibinfo{author}{\bibfnamefont{A.~R.} \bibnamefont{Atilgan}},
  \bibnamefont{and} \bibinfo{author}{\bibfnamefont{B.}~\bibnamefont{Erman}},
  \bibinfo{journal}{Fold. \& Des.} \textbf{\bibinfo{volume}{2}},
  \bibinfo{pages}{173} (\bibinfo{year}{1997}).



\bibitem[{\citenamefont{Garcia}(1992)}]{garcia92}
\bibinfo{author}{\bibfnamefont{A.}~\bibnamefont{Garcia}},
  \bibinfo{journal}{Phys. Rev. Lett.} \textbf{\bibinfo{volume}{68}},
  \bibinfo{pages}{2696} (\bibinfo{year}{1992}).

\bibitem[{\citenamefont{Amadei et~al.}(1993)\citenamefont{Amadei, Linseen, and
  Berendsen}}]{Amadei.PROT.1993}
\bibinfo{author}{\bibfnamefont{A.}~\bibnamefont{Amadei}},
  \bibinfo{author}{\bibfnamefont{A.~B.~M.} \bibnamefont{Linseen}},
  \bibnamefont{and} \bibinfo{author}{\bibfnamefont{H.~J.~C.}
  \bibnamefont{Berendsen}}, \bibinfo{journal}{Proteins}
  \textbf{\bibinfo{volume}{17}}, \bibinfo{pages}{412} (\bibinfo{year}{1993}).

\bibitem[{\citenamefont{Lindahl et~al.}(2001)\citenamefont{Lindahl, Hess, and
  {van~der~Spoel}}}]{GROMACS_3.0_01}
\bibinfo{author}{\bibfnamefont{E.}~\bibnamefont{Lindahl}},
  \bibinfo{author}{\bibfnamefont{B.}~\bibnamefont{Hess}}, \bibnamefont{and}
  \bibinfo{author}{\bibfnamefont{D.}~\bibnamefont{{van~der~Spoel}}},
  \bibinfo{journal}{J. Mol. Model.} \textbf{\bibinfo{volume}{7}},
  \bibinfo{pages}{306} (\bibinfo{year}{2001}).

\bibitem[{\citenamefont{Press et~al.}(1999)\citenamefont{Press, Teukolsky,
  Vetterling, and Flannery}}]{NR}
\bibinfo{author}{\bibfnamefont{W.~H.} \bibnamefont{Press}}  {\em et al.}, 
   \emph{\bibinfo{title}{Numerical Recipes}}
  (\bibinfo{publisher}{CUP}, \bibinfo{address}{Cambridge},
  \bibinfo{year}{1999}).

\bibitem[{\citenamefont{Hinsen et~al.}(2000)\citenamefont{Hinsen, Petrescu,
  Dellerue, Bellissent-Funel, and Kneller}}]{Hinsen.CHEMP.2000}
\bibinfo{author}{\bibfnamefont{K.}~\bibnamefont{Hinsen}}  {\em et al.}, 
   \bibinfo{journal}{Chem. Phys.}
  \textbf{\bibinfo{volume}{261}}, \bibinfo{pages}{25} (\bibinfo{year}{2000}).

\bibitem[{\citenamefont{Pitard and Orland}(1998)}]{Pitard.EurPhysLett.1998}
\bibinfo{author}{\bibfnamefont{E.}~\bibnamefont{Pitard}} \bibnamefont{and}
  \bibinfo{author}{\bibfnamefont{H.}~\bibnamefont{Orland}},
  \bibinfo{journal}{Europhys. Lett.} \textbf{\bibinfo{volume}{41}},
  \bibinfo{pages}{467} (\bibinfo{year}{1998}).

\bibitem[{\citenamefont{Daura et~al.}(2000)\citenamefont{Daura, Gademann,
  Jaun, Seebach and van Gunsteren}}]{Daura.ANGCHEM.1999}
\bibinfo{author}{\bibfnamefont{X.}~\bibnamefont{Daura}}  {\em et al.}, 
   \bibinfo{journal}{Angew. Chem. Int. Ed.}
  \textbf{\bibinfo{volume}{38}}, \bibinfo{pages}{236} (\bibinfo{year}{1999}).

\end{thebibliography}

 \end{document}